\documentstyle[12pt,aasms4]{article}

\lefthead{Funato et al.}
\righthead{Time-Symmetrized KS Hermite Scheme}

\begin{document}

%title
\title{Time-Symmetrized Kustaanheimo-Stiefel Regularization} 
%author
\author{Yoko Funato\altaffilmark{1}\altaffiltext{1}{JSPS Fellow for Japanese Junior Scientists}}
\affil{Department of Earth Science and Astronomy,\\
College of Arts and Sciences, University of Tokyo}
\authoraddr{Komaba, Meguro-ku, Tokyo, 153, Japan,
{\tt funato@chianti.c.u-tokyo.ac.jp}}
\author{Piet Hut}
\affil{Institute for Advanced Study}
\authoraddr{Princeton, NJ 08540,
{\tt piet@guinness.ias.edu}}
\author{Steve McMillan}
\affil{Department of Physics and Atmospheric Science,\\
Drexel University}
\authoraddr{Philadelphia, PA 19104,
{\tt steve@zonker.drexel.edu}}
\and
\author{Junichiro Makino}
\affil{Department of Graphics and Iniformation Sciences,\\
College of Arts and Sciences, University of Tokyo}
\authoraddr{Komaba, Meguro-ku, Tokyo, 153, Japan,{\tt makino@chianti.c.u-tokyo.ac.jp}}

\begin{abstract}
In this paper we describe a new algorithm for the long-term numerical
integration of the two-body problem, in which two particles interact
under a Newtonian gravitational potential.  Although analytical
solutions exist in the unperturbed and weakly perturbed cases,
numerical integration is necessary in situations where the
perturbation is relatively strong.  Kustaanheimo--Stiefel (KS)
regularization is widely used to remove the singularity in the
equations of motion, making it possible to integrate orbits having
very high eccentricity.  However, even with KS regularization,
long-term integration is difficult, simply because the required
accuracy is usually very high. We present a new time-integration
algorithm which has no secular error in either the binding energy or
the eccentricity, while allowing variable stepsize. The basic approach
is to take a time-symmetric algorithm, then apply an implicit
criterion for the stepsize to ensure strict time reversibility. We
describe the algorithm in detail and present the results of numerical
tests involving long-term integration of binaries and hierarchical
triples.  In all cases studied, we found no systematic error in either
the energy or the angular momentum. We also found that its calculation
cost does not become higher than those of existing algorithms.  By
contrast, the stabilization technique, which has been widely used in
the field of collisional stellar dynamics, conserves energy very well
but does not conserve angular momentum.
\end{abstract}


\section{Introduction}

The long-term numerical integration of the two-body problem is
important in several areas of astrophysics, particularly in the
context of the evolution of star clusters.  For example, binaries play
a crucial role in the secular evolution of globular clusters (see,
e.g. Hut et al. PASP 1992 and references therein). Binary evolution is
also important from the observational point of view, since many
interesting objects in globular clusters, such as X-ray sources,
millisecond pulsars, high-velocity stars, and blue stragglers, are
believed to be the result of binary interactions.  In order to study
the evolution of binaries in globular clusters, self-consistent
$N$-body simulation is a most useful tool.  In such simulations, we
integrate numerically the orbits of both the stars and the binaries in
the cluster.

There are, however, technical difficulties in obtaining useful results
from numerical experiments.  If we want to study the evolution of
globular clusters, we have to follow the orbits of tightly-bound
binaries under weak perturbations for, say, $10^{10}$ years,
corresponding to several cluster half-mass relaxation times.  Since
the period of a binary is typically less than one year, this time
interval corresponds to more than $10^{10}$ binary orbits, or $\sim
10^{13}-10^{14}$ numerical integration steps.  The computational
requirements of such an undertaking would be prohibitive; in addition,
both truncation and round-off error would be unacceptably large.  Of
course, there are several techniques used in order to reduce the
number of steps.  First of all, in practice, we deal with this
difficulty by regarding binaries as ``unperturbed'' if the relative
perturbation is less than, say, $10^{-6}$ or so.  
By this treatment, the relative calculation cost of integrating of
binary orbits greatly reduced (Makino and Hut, 1990).  The orbits of
binaries with relative perturbation larger than $10^{-6}$ must be
integrated numerically.  To reduce the number of steps in integrating
the orbits of such binaries, another technique ---
Kustaanheimo--Stiefel regularization --- has been used.

Over the last 20 years, the program NBODY5 developed by Aarseth (1963,
1985, 1994) has been the most powerful tool for performing $N$-body
simulations of star clusters. One important reason for the success of
this program is that it can handle very efficiently the formation and
evolution of binaries in $N$-body systems.  A key part of NBODY5 is
Kustaanheimo--Stiefel (KS) regularization (Kustaanheimo and Stiefel
1965), which is an extension of Levi-Civita's regularization of the
planar Keplerian problem to three dimensions (Levi-Civita, 1956).  In
KS regularization, a Kepler orbit is transformed into a harmonic
oscillator and the number of steps needed for the integration of an
orbit is reduced significantly.

In order to apply KS regularization, one first identifies a close pair
of particles using some heuristic criterion, then integrates their
center-of-mass and relative motion separately. The center-of-mass
motion is integrated in the same way as the motion of all other
particles in the system; the relative motion of the binary components
is integrated using regularized time and coordinates.

To reduce the accumulation of integration error, a technique known as
energy stabilization has been used (Baumgarte, 1973; Aarseth, 1985).
The basic idea of energy stabilization is to introduce an artificial
stabilizing term into the equation of motion so that the energy
converges to the ``true'' value. In the case of an isolated binary,
the energy is constant, and the force is simply adjusted so that
the calculated energy converges to the original value. For a perturbed
binary, the energy is integrated separately, and the force is
adjusted so that the calculated energy converges to the integrated
quantity.  The stabilization technique vastly improves energy
conservation; however, it does not conserve angular momentum. Thus,
stabilization may give the wrong answer if the evolution of the binary
eccentricity is important.

Recently, numerical integrators which have no secular error in the
integrals of motions have attracted the interests of both
astrophysicists and numerical analysts.  The integrators which had
been most extensively studied are the ``symplectic'' integrators (for
an overview, see Sanz--Serna and Calvo, 1994), which have the property
of being symplectic (or canonical) transformations in phase
space. Since a canonical transformation is one way to define a
Hamiltonian system, it is reasonable to suppose that an integration
scheme which is itself a canonical transformation might describe the
nature of the system better than an integrator which is not.  In fact,
symplectic integrators have the remarkable property that the errors in
the integrals of the motion have no secular term.  In the case of the
Kepler problem, neither the energy nor the angular momentum change
systematically in time.

Time-symmetric integrators have not received as much attention as
symplectic integrators. However, it has been demonstrated that some
time-symmetric integrators also exhibit no secular errors, regardless
of whether or not they are symplectic (Quinlan and Tremaine, 1990).
One serious problem with both symplectic and time-symmetric schemes is
that they work well only with constant stepsize. If variable stepsize
is used, the performance of a symplectic integrator degrades to that
of more ``usual'' schemes, in that it no longer automatically
conserves the integrals of motion (Skeel and Gear, 1992; Sanz--Serna
and Calvo, 1994).

Recently, Hut et al. (Hut, Makino and McMillan, 1995, hereafter HMM)
have developed a novel technique to avoid the accumulation of the
error in variable stepsize time-symmetric schemes.  They found that
the increase of the numerical error in the energy is dramatically
reduced if the size of the timestep is determined in a time-symmetric
way.  When constant stepsize is used, a time-symmetric scheme is
time-reversible, in the sense that, if we integrate the system forward
for one timestep and then integrate it backward, the system returns
precisely to the initial state.  In the case of variable stepsize,
however, the system generally does not return to the initial state,
since the stepsizes used for the forward and backward steps are
different.  HMM used a time-symmetric criterion for the timestep,
ensuring that the stepsize for the forward and backward steps are
exactly the same. They applied this scheme to the numerical
integration of the Kepler problem in physical coordinates, and found
that there was no secular error in the binding energy.

In this paper, we describe the implementation of this time-symmetric,
variable step scheme to an $N$-body code using the KS regularization.
We have developed a fourth-order integration scheme and performed a
series of numerical simulations.  We find that our algorithm has the
good properties of both symmetrized timesteps and KS regularization.
First, like all symmetrized schemes, it shows no secular error in
either energy or angular momentum.  Second, like the code with KS
regularization, it can handle binaries of arbitrary eccentricity with
a constant number of timesteps per orbit.

In section 2, we show the equations of motion and KS regularization.
In section 3, we describe the integration algorithms used.  In section
4, we describe the implementation in more detail.  In section 5 we
compare the results of our numerical experiments with those of other
schemes, considering both isolated and perturbed binaries.  As an
example of a perturbed system, we consider a hierarchical three-body
system in which the inner binary is integrated in KS coordinates and
the outer binary is integrated in Cartesian physical coordinates.
A summary and discussion are presented in section 6.

\section{Equations of Motion}

\subsection{Cartesian Coordinates}

The equations of motions of the two particles to which we apply the KS
transformation are expressed as
\begin{eqnarray}
\frac{d{\bf r}_i}{dt}  & = & {\bf v}_i,\\
m_i\frac{d{\bf v}_i}{dt} & = &
-\frac{m_im_j({\bf r}_j - {\bf r}_i)}{|{\bf r}_j - {\bf r}_i|^3}
+ {\bf F}_i, \label{eqn: 2-1}\\
(i,j & = & 1,2),\nonumber
\end{eqnarray}
where $t$ is physical time and the gravitational constant $G$ is set
to unity.  The 3-dimensional vectors ${\bf r}_i $ and ${\bf v}_i$ are
the positions and velocities of the particle $i$ in Cartesian
coordinates, $m_i $ is the mass of particle $i$, and ${\bf F}_i $ is
the force on particle $i$ due to the rest of the system.

In order to apply the KS transformation, we introduce the motion of
the center-of-mass of two particles:
\begin{eqnarray}
{\bf r}_{cm} & = & \frac{m_1{\bf r}_1 + m_2{\bf r}_2}{m_1+m_2},\\
{\bf v}_{cm} & = & \frac{m_1{\bf v}_1 + m_2{\bf v}_2}{m_1+m_2},
\end{eqnarray}
and the relative motion:
\begin{eqnarray}
{\bf r} & = & {\bf r}_2 - {\bf r}_1,\\
\frac{d{\bf r}}{dt} & \equiv & {\bf v} = {\bf v}_2 - {\bf v}_1.
\end{eqnarray}
The accelerations are given by:
\begin{eqnarray}
\frac{d{\bf v}}{dt} & = & - \frac{(m_1 + m_2){\bf r}}{|{\bf r}|^3} +
{\bf P},\label{eqn:base1}\\
\frac{d{\bf v}_{cm}}{dt} & = &  - \frac
{{\bf F}_1 + {\bf F}_2}{m_1 + m_2},
\end{eqnarray}
where the second term ${\bf P}$ in equation (\ref{eqn:base1}) is the
perturbation:
\begin{eqnarray}
{\bf P}  =  \frac{{\bf F}_2}{m_2} - \frac{{\bf F}_1}{m_1}.
\end{eqnarray}

\subsection{KS regularization}

In KS regularization, the 3-dimensional position in Cartesian
coordinates is transformed into a 4-dimensional position in KS
coordinates.  The physical time $t$ is also transformed into the KS
time $\tau $.  The transformation is expressed as follows:
\begin{eqnarray}
r & = & |{\bf r}| = u_1^2 + u_2^2 + u_3^2 + u_4^2,\\
{\bf r} & = & {\cal L}({\bf u}){\bf u} ,
\end{eqnarray}
\begin{eqnarray}
{\cal L} =
\left(
\begin{array}{rrrr}
u_1 & -u_2 & -u_3 &  u_4\\
u_2 &  u_1 & -u_4 & -u_3\\
u_3 &  u_4 &  u_1 &  u_2
\end{array}
\right)
\end{eqnarray}
\begin{eqnarray}
\frac{dt}{d\tau} = r = ({\bf u},{\bf u}),
\end{eqnarray}
where $({\bf u},{\bf u})$ is defined as the inner product of ${\bf u}$
and ${\bf u}$.
The resulting regularized equations of motion take the form
\begin{eqnarray}
{\bf u}'' & = & \frac{1}{2}h{\bf u} + \frac{1}{2}r{\cal L}^T
({\bf u}){\bf  P},
\label{eqn:2-2}
\end{eqnarray}
where $'$ denotes differentiation with respect to $\tau$, i.e. $'
\equiv d/d\tau $, superscript $\scriptstyle T$ denotes matrix
transposition, and $h$ is the specific binding energy of the two
bodies, which may be expressed in the KS system as follows:
\begin{eqnarray}
h = \frac{2{\bf u}'\cdot {\bf u}' - (m_1 + m_2)}{r}.
\label{eqn:energy1}
\end{eqnarray}
The time variation of $h$ is described as:
\begin{eqnarray}
h'  =  2{\bf u}'\cdot{\cal L}^T({\bf u}){\bf P}.
\label{eqn:energy2}
\end{eqnarray}
For an isolated binary, ${\bf P} = 0$, so $h$ is constant.

\section{Integration Techniques}

\subsection{The Hermite Scheme}

The Hermite scheme (Makino, 1990) is a fourth order
predictor-corrector scheme, which may be expressed as follows:
\begin{eqnarray}
{\bf r}_e & = & {\bf r}_b + \frac{1}{2}({\bf v}_e + {\bf v}_b) \Delta t
- \frac{1}{10}({\bf a}_e - {\bf a}_b)\Delta t^2 + 
\frac{1}{120}({\bf j}_e + {\bf j}_b)\Delta t^3,\label{eqn:her1}\\
{\bf v}_e & = & {\bf v}_b + \frac{1}{2}({\bf a}_e + {\bf a}_b) \Delta t
- \frac{1}{12}({\bf j}_e - {\bf j}_b)\Delta t^2,\label{eqn:her2}
\end{eqnarray}
where $a_X$ and $j_X$ are the acceleration and the ``jerk''
(time-derivative of the acceleration), and subscripts $b$ and $e$ denote
the values at the beginning and the end of the step, respectively.

As can be seen from equations (\ref{eqn:her1}) and (\ref{eqn:her2}),
the Hermite scheme is an implicit scheme.  The integration of one step
is carried out as follows.

\noindent
[1] Predict the positions and velocities:
\begin{eqnarray}
{\bf r}_{pred} & = & {\bf r}_b + {\bf v}_b \Delta t + \frac{1}{2}
{\bf a}_b \Delta t^2 + \frac{1}{6}{\bf j}_b \Delta t^3,\\
{\bf v}_{pred} & = & {\bf v}_b + {\bf a}_b \Delta t + \frac{1}{2}
{\bf j}_b \Delta t^2.
\end{eqnarray}

\noindent
[2] Evaluate the acceleration ${\bf a}_e$ and the jerk ${\bf j}_e$
at the end of the step.

\noindent
[3] Correct positions and velocities using equations (\ref{eqn:her1}) and 
(\ref{eqn:her2}), using the following formula:
\begin{eqnarray}
{\bf r}_e & = & {\bf r}_{pred} + \frac{1}{24}{\bf r}_b^{(4)}\Delta t^4
+ \frac{1}{120}{\bf r}_b^{(5)}\Delta t^5,\label{eqn:cor1}\\
{\bf v}_e & = & {\bf v}_{pred} + \frac{1}{6}{\bf r}_b^{(4)}\Delta t^3
+ \frac{1}{24}{\bf r}_b^{(5)}\Delta t^4,\label{eqn:cor2}
\end{eqnarray}
where ${\bf r}_b^{(i)}$ are the $i$-th derivatives which are calculated
as:
\begin{eqnarray}
{\bf r}_b^{(4)} & = & \frac{1}{\Delta t^2}\left[
-6({\bf a}_b - {\bf a}_e) - \Delta t (4{\bf j}_b + 2{\bf j}_e)\right],
\label{eqn:cor3}\\
{\bf r}_b^{(5)} & = & \frac{1}{\Delta t^3}\left[
12({\bf a}_b - {\bf a}_e) + 6 \Delta t ({\bf j}_b + {\bf
j}_e)\right]\label{eqn:cor4}
\end{eqnarray}

\noindent
[4] Repeat [2]--[3] until the values of ${\bf r}_e$ and ${\bf v}_e$
converge.

Equations (\ref{eqn:cor1})--(\ref{eqn:cor4}) are equivalent to equations 
(\ref{eqn:her1}) and (\ref{eqn:her2}). However, if
the stepsize is calculated from higher order derivatives, 
equations (\ref{eqn:cor1})--(\ref{eqn:cor4}) are preferred.

The local truncation error of equation (\ref{eqn:her1}) is $O(\Delta t^6)$
and that of equation (\ref{eqn:her2}) is $O(\Delta t^5)$. The global 
error is $O(\Delta t^4)$ for both $r$ and $v$. 
One could also use the truncated form, 
\begin{eqnarray}
{\bf r}_e & = & {\bf r}_b + \frac{1}{2}({\bf v}_e + {\bf v}_b) \Delta
t - \frac{1}{12}({\bf a}_e - {\bf a}_b)(\Delta t)^2,
\label{eqn:her3}
\end{eqnarray}
which has a local error of $O(\Delta t^5)$ and a global error (that
is, integrated over some fixed time interval) of $O(\Delta t^4)$ for
both $r$ and $v$.  For the truncated Hermite scheme, the local
truncated error in the position during one step is $O(\Delta t^5)$,
while it is $O(\Delta t^6)$ in the full Hermite scheme.  In either
case, the global energy error is $O(\Delta t^4)$.

\subsection{Time-symmetrization}

We now describe the basic idea of the time-symmetric scheme.  First,
we consider an integration scheme with a constant stepsize.  An
arbitrary integration scheme with constant stepsize $\Delta t$ may be
expressed compactly as follows:
\begin{eqnarray}
\xi_e =
f(\xi_b, \xi_e, \Delta t),
\end{eqnarray}
where $\xi$ represents the phase-space variables, i.e. $\xi = ({\bf
r},{\bf v})$.  If $f$ does not depends on $\xi_e$, the scheme is
explicit. Otherwise, the scheme is implicit.

If the scheme is time-symmetric, the following equation holds
\begin{eqnarray}
\bar{\xi } \equiv f(\xi_e, \xi_b, -\Delta t) = \xi_b.
\end{eqnarray}
For example, a leap-frog scheme with constant stepsize, which is a
well-known symplectic scheme, is also time-symmetric.  The leap-frog
scheme is an explicit scheme. However, many useful time-symmetric
schemes are implicit.

The simplest example of an implicit time-symmetric scheme is the
implicit trapezoidal rule, which may be expressed as:
\begin{eqnarray}
\xi_e = \xi_b + \frac{1}{2}[\phi(\xi_b) + \phi(\xi_e)] \Delta t,
\label{eqn:trapezoidal}
\end{eqnarray}
where 
\begin{eqnarray}
\phi(\xi) \equiv \frac{d\xi}{dt}.
\end{eqnarray}
In order to integrate for one step, the implicit equation
(\ref{eqn:trapezoidal}) is solved iteratively until convergence is
obtained.  There are also time-symmetric integrators of higher order.
The Hermite scheme [equations (\ref{eqn:her1}) and (\ref{eqn:her2}),
or (\ref{eqn:her3}) and (\ref{eqn:her2})] is an example of a
fourth-order symmetric integrator.

Now we consider time-symmetric schemes with variable stepsize.  If the
scheme is time-symmetric, the following equations are satisfied:
\begin{eqnarray}
\begin{array}{lcl}
\xi_e & = & f(\xi_b, \xi_e, \Delta t_b),\\
\xi_b & = & f(\xi_e, \xi_b, -\Delta t_e),
\label{eqn:symm1}
\end{array}
\end{eqnarray}
where $\Delta t_b$ and $\Delta t_e $ are the stepsize determined at
the beginning of the step and the end of the step by some function
$\delta(\xi)$, i.e.
\begin{eqnarray}
\Delta t_b  =  \delta(\xi_b, \xi_e);\,\,\,
\Delta t_e  =  \delta(\xi_e, \xi_b).
\end{eqnarray}
In order to guarantee that equations (\ref{eqn:symm1}) hold, it is
sufficient to make the stepsize criterion time-symmetric:
\begin{eqnarray}
\delta(\xi_b, \xi_e)  =  \delta(\xi_e, \xi_b).
\label{eqn:sym}
\end{eqnarray}
One simple way to construct a timestep criterion that satisfies
equation (\ref{eqn:sym}) is to take the average of stepsize calculated
at the beginning and the end of the step. For example, 
\begin{eqnarray}
\delta(\xi_b, \xi_e)  \equiv  \frac{1}{2}[s(\xi_b) + s(\xi_e)],
\label{eqn:symstep1}
\end{eqnarray}
or
\begin{eqnarray}
\delta(\xi_b, \xi_e)  
\equiv  \sqrt{\frac{1}{2}[s(\xi_b)^2 + s(\xi_e)^2]}.
\label{eqn:symstep2}
\end{eqnarray}
Equations (\ref{eqn:symstep1}) and (\ref{eqn:symstep2}) are implicit
equations.  The stepsize is determined by solving equation
(\ref{eqn:symstep1}) or (\ref{eqn:symstep2}) coupled with equations
(\ref{eqn:her1}) and (\ref{eqn:her2}) [or (\ref{eqn:her3}) and
(\ref{eqn:her2})].

\subsection{Energy Stabilization}

Energy stabilization has been used in
traditional programs to conserve the binding energy of a
binary.  In a stabilized scheme, energy conservation is enforced by
introducing an artificial acceleration into the  equation of
motion, as follows (Baumgarte 1973):
\begin{eqnarray}
\frac{d^2{\bf r}}{dt^2} = \frac{(m_1+m_2){\bf r}}{|{\bf r}|^3}
+ {\bf P} - \left(\frac{\alpha}{\Delta t}\right)\frac{(h_{integ} - h_{direct}){\bf
v}} {({\bf v}\cdot {\bf v})}.\label{eqn:stab1}
\end{eqnarray}
Following the prescription by Aarseth (1985), equation
(\ref{eqn:stab1}) is transformed to KS coordinates as:
\begin{eqnarray}
\frac{d^2{\bf u}}{d\tau^2} = \frac{1}{2}h{\bf u} +
\frac{1}{2}r{\cal L}^T{\bf P} - 
\left(\frac{\alpha}{\Delta \tau}\right) \frac{ r (h_{integ} - h_{direct}){\bf u}'}{M}.
\label{eqn:stab2}
\end{eqnarray}
The third term of the right-hand side of equation (\ref{eqn:stab2}) is
the artificial stabilizing acceleration. The terms $h_{integ} $ and
$h_{direct}$ are the specific binding energy of the binary obtained by
integrating equation (\ref{eqn:energy2}) and evaluating equation
(\ref{eqn:energy1}), respectively. 
Figure \ref{fig0} shows the time variation of $h_{direct}$ for different values
of $\alpha$.  When $\alpha $ is less than 0.3, the behavior of the
energy is simple.  When $\alpha $ is larger than 0.3, the behavior of
the energy is complex.  With small $\alpha $, $h_{direct}$ cannot
follow $h_{integ}$ fast enough.  In fact, for $\alpha = 0.1$, a
secular error of $\sim 5\times 10^{-7}$/orbit was observed. For large
$\alpha $, the spikes around the periastron is very large and the
error is enough adjusted every period.  Aarseth recommended to use
$\alpha = 0.4$.  In this paper, we set $\alpha$ to 0.4.

\section{Implementation of the Symmetrized KS Hermite Scheme}

\def\dtau#1#2{ #2^{(#1)}}

\subsection{The KS Hermite scheme}

The equations of motion of the KS binary are (14) and (16).  We
integrate them using the time-symmetrized Hermite scheme.
\begin{eqnarray}
{\bf u}_e & = & {\bf u}_b + \frac{1}{2}(\dtau{1}{{\bf u}}_e + \dtau{1}{{\bf u}}_b) \Delta \tau
- \frac{1}{10}(\dtau{2}{{\bf u}}_e - \dtau{2}{{\bf u}}_b)\Delta \tau^2 + 
\frac{1}{120}(\dtau{3}{{\bf u}}_e + \dtau{3}{{\bf u}}_b)\Delta \tau^3,\label{eqn:ksher1}\\
\dtau{1}{{\bf u}}_e & = & \dtau{1}{{\bf u}}_b + \frac{1}{2}(\dtau{2}{{\bf u}}_e + \dtau{2}{{\bf u}}_b) 
\Delta \tau
- \frac{1}{12}(\dtau{3}{{\bf u}}_e - \dtau{3}{{\bf u}}_b)\Delta \tau^2,\label{eqn:ksher2}\\
h_e & = & h_b + \frac{1}{2}(\dtau{1}h_0 + \dtau{1}h_1)\Delta \tau
- \frac{1}{12}(\dtau{2}h_1 - \dtau{2}h_0)\Delta \tau^2.
\end{eqnarray}
Here and below we denote the $i$-th derivative with respect to a variable
$\tau$ by using the operator $^{(i)}$. For example,
\begin{eqnarray}
\frac{d^2{\bf u}}{d\tau^2}  &=&  {\bf u}^{(2)}.\\
\end{eqnarray}
The second and third derivatives are calculated as (Aarseth
1985):
\begin{eqnarray}
\dtau{2}{{\bf u}}_b & = & \frac{1}{2}h_b{\bf u}_b +
\frac{1}{2}r_b{\cal L}_b^T({\bf u}_b){\bf P}_b,\\
\dtau{3}{{\bf u}}_b & = & \frac{1}{2}h_b\dtau{1}{{\bf u}}_b + \frac{1}{2}\dtau{1}h_b{\bf u}_b
+ \frac{1}{2}\frac{d}{d\tau}[r_b{\cal L}_b^T({\bf u}_b){\bf P}_b]\nonumber\\
& = & \frac{1}{2}h_b\dtau{1}{{\bf u}}_b + \frac{1}{2}\dtau{1}h_b{\bf u}_b
+ \frac{1}{2}\frac{d}{d\tau}[r_b{\cal L}_b^T({\bf u}_b)]{\bf  P}_b
+ \frac{1}{2}r_b{\cal L}_b^T({\bf u}_b)r_b{\bf  J}_{p,b},\\
\dtau{1}h & = & - 2\dtau{1}{{\bf u}}\cdot{\cal L}^T({\bf u}){\bf P},\\
\dtau{2}h & = & - 2[\dtau{2}{{\bf u}}\cdot{\cal L}^T({\bf u}){\bf P} + 
\dtau{1}{{\bf u}}\cdot{\cal L}^T(\dtau{1}{{\bf u}}){\bf P} + 
\dtau{1}{{\bf u}}\cdot{\cal L}^T({\bf u})r{\bf J}_p]
\end{eqnarray}
The quantities $R_b$ and ${\cal L}_b$ are, respectively, the
separation of the binary in 
Cartesian coordinates and the transformation matrix at time $\tau =
\tau_b$.  Because of the numerical difficulties associated with 
the singularity in equation (\ref{eqn:energy1}) as $R\rightarrow 0$
(Aarseth, 1972),
we integrate the binding energy $h$ of the binary instead of
calculating it from its definition at each step.

\subsection{Timestep Function}

The timestep symmetrization technique described in section 3.2 can be
applied to an arbitrary algorithm for calculating the stepsize.  

The standard Aarseth formula (Aarseth 1985), which 
is being used in most $N$-body programs, is expressed as
\begin{eqnarray}
\Delta \tau = s({\bf u}) = \sqrt{\eta
\frac{|\dtau{4}{{\bf u}}||\dtau{2}{{\bf u}}| + |\dtau{3}{{\bf u}}|^2}
     {|\dtau{5}{{\bf u}}||\dtau{3}{{\bf u}}| + |\dtau{4}{{\bf u}}|^2}}.
\label{eqn:stepsize1}
\end{eqnarray}
Here $\eta $ is an accuracy parameter.
The stepsize criterion is known to have good properties. 

For general $N$-body problem, this criterion works very well.  For a
perturbed two-body problem with KS regularization, however, a more
simpler schemes might be sufficient. We tried two formula.

If we adopt the symmetrized scheme, we may be able to carry out an
accurate and fast time integration without the standard formula. We
can use simpler functions such as
\begin{eqnarray}
\Delta \tau = \eta \frac{|{\bf u}|}{|{\bf u}^{(1)}|},
\label{eqn:stepsize2}
\end{eqnarray} 
or
\begin{eqnarray}
\Delta \tau = s({\bf u}) = \sqrt{\eta
\frac{|\dtau{2}{{\bf u}}||{\bf u}| + |\dtau{1}{{\bf u}}|^2}
     {|\dtau{3}{{\bf u}}||\dtau{1}{{\bf u}}| + |\dtau{2}{{\bf u}}|^2}},
\label{eqn:stepsize3}
\end{eqnarray} 
as the stepsize function for the KS timestep $\Delta\tau$.

Here ``simpler'' means that equations (\ref{eqn:stepsize2}) and
(\ref{eqn:stepsize3}) don't require the 
higher-order terms 
such as
$\dtau{4}{{\bf u}}$ and $\dtau{5}{{\bf u}}$.  It 
makes the convergence faster.

Note that each of the functions (\ref{eqn:stepsize1}) and
(\ref{eqn:stepsize3}) gives a constant stepsize in KS coordinates for
the time integration of a simple binary.  
For a time-symmetrized scheme, it is better to use the function which
gives a nearly constant stepsize.  This is because the estimation of
the initial value of iterations becomes easier.

\subsection{Implementation in General $N$-body Systems}

\subsubsection{For a Few Body Problem}

We now describe the implementation of the symmetrized KS Hermite
scheme in a full $N$-body code.  We consider here only a
shared-timestep scheme, in which the whole system is advanced with a
single global timestep.  This method is sufficient for 3- or 4-body
systems. 

The procedure for a single step of the time integration is as follows.
At the beginning of the step, the relative position and velocity of
the binary components in KS coordinates are known. The positions and
velocities of other particles are given in Cartesian coordinates.  The
binary perturbation ${\bf P}$ and its time derivative ${\bf J}_p$ are
first calculated in Cartesian coordinates.  Then the relative position
of the binary components is predicted in KS coordinates with KS time
step $\Delta\tau$, while the center of the mass of binary and all
other particles are predicted in Cartesian coordinates with physical
time step $\Delta t$, which is a function of $\Delta\tau$ and ${\bf
u}$.

The stepsize $\Delta\tau$ in the regularized system and the
corresponding physical stepsize $\Delta t$ must be calculated in such
a way as to ensure the time-reversibility of the whole $N$-body
system.  We use the following formula to advance the physical time:
\begin{eqnarray}
t_1 & = & t_0 + \Delta t,\\
\Delta t  & = & T(\Delta \tau)
= \dtau{1}t_{\frac{1}{2}}\Delta \tau
+ \dtau{3}t_{\frac{1}{2}}\frac{\Delta \tau^3}{24}
+ \dtau{5}t_{\frac{1}{2}}\frac{\Delta \tau^5}{1920}.
\label{eqn:time1}
\end{eqnarray}
Equation (\ref{eqn:time1}), which is a 5th order expression, is
calculated by using the first through the fifth derivatives of the
positions (i.e. all derivatives available during the integration).
The formulation is the same as used by Aarseth (1985), except that we
use the derivatives at the midpoint of the integration interval, which
eliminates all even derivatives.  Here $\dtau{i}t_{\frac{1}{2}}$ are
the $i$-th derivatives at $\tau =
\tau_0 + 0.5 \Delta \tau $ in KS time.  They are calculated as:

\begin{eqnarray}
\dtau{1}t_{\frac{1}{2}} & = &
r = {\bf u}\cdot {\bf u},\\
\dtau{3}t_{\frac{1}{2}} & = & 2({\bf u}\cdot \dtau{2}{{\bf u}}
+ \dtau{1}{{\bf u}}\cdot \dtau{1}{{\bf u}}),\\
\dtau{5}t_{\frac{1}{2}} & = & 2[{\bf u}\cdot \dtau{4}{{\bf u}} +
                      4(\dtau{1}{{\bf u}}\cdot \dtau{3}{{\bf u}}) +
                      3(\dtau{2}{{\bf u}}\cdot \dtau{2}{{\bf u}})],
\end{eqnarray}
where
\begin{eqnarray}
{\bf u}_{\frac{1}{2}} & = &
{\bf u}_0 + \frac{\Delta \tau}{2}\dtau{1}{{\bf u}}_0 + \frac{\Delta \tau}{8}\dtau{2}{{\bf u}}_0
+ \frac{\Delta \tau^3}{48}\dtau{3}{{\bf u}}_0
+ \frac{\Delta \tau^4}{384}\dtau{4}{{\bf u}}_0
+ \frac{\Delta \tau^5}{3840}\dtau{5}{{\bf u}}_0,\\
\dtau{1}{{\bf u}}_{\frac{1}{2}} & = &
\dtau{1}{{\bf u}}_0 + \frac{\Delta \tau}{2}\dtau{2}{{\bf u}}_0
+ \frac{\Delta \tau^2}{8}\dtau{3}{{\bf u}}_0
+ \frac{\Delta\tau^3}{48}\dtau{4}{{\bf u}}_0
+ \frac{\Delta \tau^4}{384}\dtau{5}{{\bf u}}_0,\\
\dtau{2}{{\bf u}}_{\frac{1}{2}} & = &
\dtau{2}{{\bf u}}_0 + \frac{\Delta \tau}{2}\dtau{3}{{\bf u}}_0
+ \frac{\Delta \tau^2}{8}\dtau{4}{{\bf u}}_0
+ \frac{\Delta \tau^3}{48}\dtau{5}{{\bf u}}_0,\\
\dtau{3}{{\bf u}}_{\frac{1}{2}} & = &
\dtau{3}{{\bf u}}_0 + \frac{\Delta \tau}{2}\dtau{4}{{\bf u}}_0
+ \frac{\Delta \tau^2}{8}\dtau{5}{{\bf u}}_0,\\
\dtau{4}{{\bf u}}_{\frac{1}{2}} & = &
\dtau{4}{{\bf u}}_0 + \dtau{5}{{\bf u}}_0\frac{\Delta \tau}{2},
\end{eqnarray}

It may be sufficient to use the 4th-order Hermite scheme (the
truncated Hermite scheme) to integrate the physical time, since the
phase-space variables are globally integrated to 4th-order
accuracy.  More detailed error analysis will be given elsewhere.

The details of a single integration step are as follows.

\noindent
[1]
Predict the positions and velocities of all particles.

\noindent
[2] Evaluate all accelerations and jerks using predicted positions and
velocities.

\noindent
[3] Calculate the fourth and fifth derivatives of the relative
position of the binary components in KS coordinates at both the
beginning and the end of the step using the following formulae:
\begin{eqnarray}
\dtau{4}{{\bf u}}_{b} & = & - \frac{1}{\Delta \tau_b^2}
\left[6(\dtau{2}{{\bf u}}_b - \dtau{2}{{\bf u}}_e) +
2\Delta \tau_b (2\dtau{3}{{\bf u}}_b + \dtau{3}{{\bf u}}_e)\right],\\
\dtau{5}{{\bf u}}_{b} & = & \frac{1}{\Delta \tau_b^3}
\left[12(\dtau{2}{{\bf u}}_b - \dtau{2}{{\bf u}}_e) +
6\Delta \tau_b (\dtau{3}{{\bf u}}_b + \dtau{3}{{\bf u}}_e)\right],\\
\dtau{4}{{\bf u}}_e & = & \dtau{3}{{\bf u}}_b +
                     \Delta \tau_b \dtau{4}{{\bf u}}_b,\\
\dtau{5}{{\bf u}}_e & = & \dtau{4}{{\bf u}}_b.
\end{eqnarray}
We calculate the higher order derivatives for other particles and the
specific energy of the binary in the same way.

\noindent
[4] Calculate the new stepsize and the physical stepsize at the
beginning of the step. For example, 
\begin{eqnarray}
\Delta \tau_{new} & = & \frac{1}{2}\left[
s({\bf u}_b) + s({\bf u}_e)\right],\\
\Delta t_{new} & = & T(\Delta \tau_{new}),
\end{eqnarray}

\noindent
[5] Correct the integrated values using the new stepsize $ \Delta
\tau_{new}$ :
\begin{eqnarray}
{\bf u}_e & = & {\bf u}_b + 
\dtau{1}{{\bf u}}_b\Delta \tau_{new} +
\frac{1}{2}\dtau{2}{{\bf u}}_b\Delta \tau_{new}^2 +
\nonumber \\ && \,\,\,\,\,\,\,\,\,
\frac{1}{6}\dtau{3}{{\bf u}}_b\Delta \tau_{new}^3 +
\frac{1}{24}\dtau{4}{{\bf u}}_b\Delta \tau_{new}^4 +
\frac{1}{120}\dtau{5}{{\bf u}}_b\Delta \tau_{new}^5,
\\
\dtau{1}{{\bf u}}_e & = & \dtau{1}{{\bf u}}_b + \dtau{2}{{\bf u}}_b\Delta \tau_{new} +
\frac{1}{2}\dtau{3}{{\bf u}}_b\Delta \tau_{new}^2 +
\frac{1}{6}\dtau{4}{{\bf u}}_b\Delta \tau_{new}^3 +
\frac{1}{24}\dtau{5}{{\bf u}}_b\Delta \tau_{new}^4.
\end{eqnarray}

\noindent
[6] Repeat procedures [2]--[5] until both the stepsize and the
integrated variables converge.

About the convergence of integrated variables, there is one simple
method in which the procedure shall be repeated until the variables of
all particles converge.  Even if we adopt this simplest method, the
number of iterations of the procedure is not so large for a few-body
systems.  In fact, in an example shown in the following section, the
number of iterations was typically 4 for the case of an integration of a
hierarchical triplet.  The number does not much exceed that required
to converge the phase space variables in the ordinary Hermite scheme,
which is $2 \sim 3$.

Even for more complicated cases, we may make the number of
iterations small if we choose an appropriate criterion formula to
determine the stepsize $\Delta \tau$, or if we choose an appropriate
accelerator to converge the variables.

\subsubsection{For a Large N Body Systems}

The outline of the procedure of integration of one step is same as
that for the cases of few body systems.  However, there are two
difficulties. One is the increase of the number of iterations until
convergence. Another is how to implement both symmetrized scheme and
individual time step scheme.
For the system with larger value of $N$, the number of iteration may
become larger.  For larger values of $N$, individual timesteps become
necessary to achieve reasonable efficiency (Aarseth 1985).

Implementation of time-symmetrization for individual timesteps will be
discussed elsewhere.
Here we only give a comment about this point.  In a time integration
of a large $N$-body systems, it is not necessary to symmetrize all
variables of all particles. To symmetrize the integration of binaries
(and its close neighbors) would be sufficient.

\section{Numerical Experiments} 

Figures \ref{fig1}a--\ref{fig1}c show the behavior of the errors in the energy,
eccentricity, and angular momentum of a binary with initial
eccentricity $e = 0.999999$.  In figures
\ref{fig1}a--\ref{fig1}c, the values at the apocenter are plotted.
These figures show that there is no secular error for 2000 binary
periods. The specific energy is calculated by equation (15). 

In the following subsections, we compare
(1) the symmetrized KS--Hermite scheme, 
(2) the ``plain'' (i.e. neither symmetrized nor stabilized) KS--Hermite scheme,
and
(3) the stabilized KS--Hermite scheme, 
for $e = 0.9$. We have carried out other comparisons, with
initial eccentricities ranging from $e = 0.0$ to $e = 0.999999$.  In
all cases we obtain qualitatively the same results as are shown
below.

We have investigated the dynamical evolution of both a simple binary and a
hierarchical triple. The initial conditions used in our experiments
are summarized in Table 1.  In the case of the binary the integration
was performed for $2000$ orbits.  For the triple, we integrated the
system for 2000 periods of the inner binary.  

We used equation (\ref{eqn:stepsize3}) as the stepsize function and
equation (\ref{eqn:symstep2}) as the symmetric criterion.  The
coefficient $\eta$ in equation (\ref{eqn:stepsize3}) was set to
0.01, corresponding to a stepsize of roughly $1/30$ of the (inner)
binary period.  The condition to stop the iterations was $|\Delta
\tau_e - \Delta \tau_b| < 1.0\times 10^{-15}$.  The total number of
timesteps was therefore $\sim 6\times 10^{4}$ in all cases.  All
calculations were done in double precision (16-digit accuracy).

\subsection{A Simple Binary}

For this case, the number of iterations required by the symmetrization
of stepsize $\Delta \tau $ is 2.  The number of iterations necessary
is small since $\Delta
\tau$ should be constant, i.e., $\Delta \tau_b = \Delta \tau_e $ for 
the criterion (\ref{eqn:stepsize3}).

Figures \ref{fig2} and \ref{fig3} show the result of an integration of
the Kepler two-body problem for $2000$ orbital periods of the binary.
Figure \ref{fig2} shows the error in the total energy (i.e. the error
in the binding energy of the binary) for the three schemes listed
above.  Here we plot the error in the energy calculated from equation
(\ref{eqn:energy1}), not the error in the energy integrated using
equation (\ref{eqn:energy2}).  The reason is that the integrated
$h_{integ}$ is constant, because the binary is unperturbed. The
calculated energy thus indicates the accuracy of the orbital
integration. We also use this calculated energy for other cases
discussed below.

Figure \ref{fig2} shows that both the symmetrized and the stabilized
schemes conserve energy, while the error in the energy increases
linearly with time for the plain integrator.  In the symmetrized case,
the energy error $\Delta E/E$ is less than $10^{-12}$ after $10^3$
orbits, or about $10^{5}$ steps, indicating that it is mostly due to
round-off error.

We have plotted the change of energy at regular time intervals. The
error of the energy becomes large at the pericenter, though the error
returns back to nearly zero at the apocenter.  As a result, the
beating of the frequency of the binary orbit and that of our sampling
is observed for the symmetrized and stabilized cases. The behavior of
energy in the stabilized case in figure \ref{fig2} is more complicated than
that in the symmetrized case, since the time variation of error around
the pericenter is complicated (see figure \ref{fig0}d).

Figure \ref{fig3} is the same as Figure \ref{fig2}, but for the error in the angular
momentum $\Delta A$.  Figure \ref{fig3} shows that both the symmetrized and the
plain schemes conserve total angular momentum, while the error in the
angular momentum increases linearly in the stabilized case.  The
relative error in the angular momentum in the symmetrized case is
again less than $10^{-12}$.

Figure \ref{fig4} is the same as Figure \ref{fig2}, but for the eccentricity.  This
figure makes explicit the fact that the eccentricity of the binary is
conserved by the symmetrized integrator, but not by either the plain
or the stabilized schemes.  This result is quite natural since the
eccentricity depends on both the energy and the angular momentum.  The
plain scheme conserves angular momentum but not energy, while the
stabilized scheme conserves energy but not angular momentum. Thus
neither scheme preserves the eccentricity.  Only the symmetrized
scheme conserves both energy and eccentricity up to round-off error.

\subsection{A Hierarchical Triple}

For this case, the number of iterations required by the symmetrization
is about 4.  That required by the convergence of the phase space
variables is $2\sim 3$ for the non-symmetrized Hermite scheme.
The symmetrized scheme, therefore, is not 
much expensive than the non-symmetrized scheme.

Figures \ref{fig5} and \ref{fig6} show, for our three integration
schemes, the relative errors in the total energy $\Delta E/E$ (Figure
\ref{fig5}) and the total angular momentum $\Delta A$ (Figure
\ref{fig6})of the three-body system.  In this case again, the relative
error in the energy using the symmetrized scheme is no more than
$10^{-12}$ and the error in the angular momentum is about $10^{-11}$
after $10^5$ steps.

Figures \ref{fig7}-\ref{fig10} show the evolution of the inner binary.
From the analytical treatment presented in the Appendix, the
time evolution of energy, angular momentum and eccentricity of the
(weakly perturbed) inner
binary may be expressed as follows. 
\begin{eqnarray}%
\Delta E/E & \cong & 0.0,\\
\Delta A & \cong & 0.5\cdot 10^{-3} [1-\cos(2\Omega_0 t)],\\
\Delta e & \cong & 0.25 \cdot 10^{-3} [-1+\cos(2\Omega_0 t)],
\end{eqnarray}
where subscript 0 indicates unperturbed values.

Figure \ref{fig7} shows the variation of the specific energy of the
binary. The specific energy is conserved in both the symmetrized and
the stabilized cases, while it increases linearly in the plain case.
Figure \ref{fig8} is the same as Figure \ref{fig7}, but for the binary
angular momentum.  Figure \ref{fig9} is the same as Figure \ref{fig7},
but for the binary eccentricity.  In figures \ref{fig7}, \ref{fig8}
and \ref{fig9}, the bare values at the apocenter of the inner binary
are plotted.  As shown in figures \ref{fig7}--\ref{fig9}, since the
amplitude of the variations during one period of the outer binary is
much larger than the numerical errors incurred within one period, as
shown in equations (64)-(66), the numerical error could not be seen if
the bare values were plotted.  In order to show the secular variation,
we applied the linear regression. Straight lines in figures
\ref{fig7}--\ref{fig9} are the results of the least 
square fitting.  Figures \ref{fig7}, \ref{fig8} and \ref{fig9} show
that there are no significant secular variations in the symmetrized
case, and that the eccentricity of the binary is not correctly
followed by the stabilizing integrator.  Comparing figures \ref{fig6}
and \ref{fig8}, we see that the error in the total angular momentum
comes mostly from the integration of the KS binary in the stabilized
case.

Figures \ref{fig10}a and b show the time variation of the angular
momentum and the eccentricity of the inner binary during the first few
periods of the outer binary.  The dotted curves correspond to the
analytical values, while the solid curve corresponds to the results of
numerical experiments in the symmetrized case.
Figures \ref{fig10}a and b show that the numerical experiments agree
well with the analytical results.

To summarize, our numerical experiments show that both
the total system and the inner binary are integrated correctly in the
symmetrized case.  In other words, the momentum transfer between the
outer particle and the inner binary is accurately followed. Thus, our
scheme follows the total 
system consistently, even though part of the system is expressed in
KS coordinates and the rest in physical coordinates.

\section{Summary and Discussion}

We have developed a new algorithm for the integration of perturbed
binary motion.  The algorithm is constructed using two techniques: KS
regularization and time-symmetrization.  Our algorithm allows variable
stepsize, but produces no secular error in either the binding energy
or the eccentricity.  We have presented the results of numerical
experiments for one case with $e = 0.9$. We have also performed
experiments for other eccentricities, and have confirmed the
superiority of the method in cases ranging from circular ($e = 0$) to
highly elongated ($e = 0.999999$).

Our scheme is composite, in the sense that the internal motion of the
binary and the motion of the rest of the system are integrated in
separate coordinate systems.  This is rather different from the usual
applications of symmetric or symplectic integrators, where all degrees
of freedom are integrated in the same way.  For an unperturbed binary,
our algorithm simply applies the symmetrization to the system with a
transformed Hamiltonian. Thus it is not surprising that our algorithm
works well for unperturbed binaries.

For perturbed systems, the result of numerical experiment shows that
the error caused by any inconsistency between the integration of the
system in the physical coordinates and the KS coordinates is smaller
than the error caused by the integration of KS binary.  Furthermore,
our numerical experiments shows that the calculation cost of the new
scheme is not very high compared with that of ordinary Hermite
scheme. This result suggests that our scheme is valid for weakly
perturbed cases.  An accurate integration of long surviving binary is
the main target of the time symmetrized scheme. Since the perturbation
on such a binary is weak, our scheme is excellent from a practical
point of view.

In conclusion, we have shown that the time-symmetrization scheme,
introduced by Hut et al. (1995) can be successfully generalized to
include KS regularization.  It is clear that our scheme can provide
excellent accuracy for individual demanding cases.  Under which
circumstances this new scheme will prove to be competitive will have
to await a detailed testing in realistic simulations.

\acknowledgements
The authors thank Sverre Aarseth for offering his
KS-Hermite program.  YF thanks Douglas Heggie for helpful discussions
and encouragement.  This work was partly supported by Grant-in-Aid for
Specially Promoted Research (04102002) of the Ministry of Education,
Science and Culture, 
partly by Grant-in-Aid for Japanese Junior
Scientists(05-3020) and partly by Joint Research Program Between the
Royal Society, the British Council and JSPS.

%
% Appendix
%

\appendix

\section{Appendix:~~Analytical treatment of a hierarchical triple}

We introduce physical coordinates as shown in figure 12. The origin is
placed at the center of mass of the inner binary.  The masses of
particles 1 and 2 are $m_1 = m_2 = m = 1/2$.  
%Hereafter we consider only the case of $m_1 = m_2$. This case is
%special and the study of the effect of higher order terms is required
%since the quadratic terms vanish.
The unperturbed motion of particles 1 and 2 may be expressed as
\begin{eqnarray}
{\bf r}_1 & = & (r_1 \cos \phi , r_1 \sin \phi),\label{eqn:a1}\\
{\bf v}_1 & = & (\dot{r}_1 \cos \phi - r_1\dot{\phi} \sin \phi,
\dot{r}_1 \sin \phi + r_1\dot{\phi} \cos \phi) , \label{eqn:a2}\\
{\bf r}_2 & = & - (r_2 \cos \phi , r_2 \sin \phi),\label{eqn:a3}\\
{\bf v}_2 & = & - (\dot{r}_2 \cos \phi - r_2\dot{\phi} \sin \phi,
\dot{r}_2 \sin \phi + r_2\dot{\phi} \cos \phi) \label{eqn:a4}, 
\end{eqnarray}
where ${\bf r}_i $ and ${\bf v}_i $ are the position and velocity and
the radius of particle $i$, $r_i = |{\bf r}_i|$, and $\phi $ is the
true anomaly of the inner binary.

The unperturbed motion of particle 3
is expressed as a circular orbit around the center of mass of the
binary:
\begin{eqnarray}
{\bf r}_3  =  (R \cos \psi , R \sin \psi) 
 =  (R \cos \Omega t , R \sin \Omega t) \label{eqn:a5},
\end{eqnarray}
where $R$ is the distance from the third body  to the binary center of
mass and $\psi $ is the true anomaly of the outer orbit.

Since $m_1 = m_2$, $r_1 = r_2 = r$, where $r$ is half the
distance between particles 1 and 2.
The accelerations of particles 1 and 2 
due to particle 3 are:
\begin{eqnarray}
{\bf a}_1 & = & m_3
\frac{(R \cos \psi - r \cos \phi , R \sin \psi - r \sin \phi)}
{[R^2 + r^2 - 2Rr\cos(\phi - \psi)]^{\frac{3}{2}}}, \label{eqn:a6}\\
{\bf a}_2 & = & m_3
\frac{(R \cos \psi + r \cos \phi , R \sin \psi + r \sin \phi)}
{[R^2 + r^2 + 2Rr\cos(\phi - \psi)]^{\frac{3}{2}}}\label{eqn:a7}.
\end{eqnarray}

For unperturbed motion of the inner binary,
the binary separation and time derivatives of the separation and the
phase angle are 
\begin{eqnarray}
r_1 & = & r_2 ~~ = ~~ r ~~ = 
~~ \frac{a(1-e^2)}{2(1 + e \cos \phi)}, \label{eqn:a8}\\
\dot{r}_1 & = & \dot{r}_2 ~~ = 
~~ \dot{r} ~~ = ~~ \frac{A e \sin \phi}{2a(1-e^2)},\label{eqn:a9}\\
\dot{\phi} & = & \frac{A}{4r^2} = 
 \frac{A(1+e\cos\phi)^2}{a^2(1-e^2)^2} \label{eqn:a10}.
\end{eqnarray}
Here $a$, $e$ and $A$ are the semi-major axis, the eccentricity, and the
specific angular momentum of the relative motion of the inner binary;
$A = \sqrt{a(1-e^2)}$.

\subsection{Energy Variation}

The specific binding energy of the inner binary, and its variation due
to infinitesimal changes $\delta {\bf x}_i$ and $\delta {\bf v}_i$, are
\begin{eqnarray}
E & = & \sum_{i=1,2}\frac{1}{2} {\bf v}_i\cdot {\bf v}_i - \frac{m}{2r}, \label{eqn:a11}\\
\delta E & = & \sum_{i=1,2}{\bf v}_i\cdot \delta{\bf v}_i + 
\frac{m {\bf r}\cdot {\bf \delta} {\bf r}}{2r^3} 
= -\sum_{i=1,2}{\bf v}_i\cdot \delta{\bf v}_i, \label{eqn:a12}
\end{eqnarray}
Expressing the variation of the velocity as $\delta
{\bf v}_i = {\bf a}_i \delta t $, we obtain
\begin{eqnarray}
\delta E & = & -\sum_{i=1,2}{\bf v}_i\cdot {\bf a}_i \delta t. \label{eqn:a13}
\end{eqnarray}

From equations (\ref{eqn:a1})--(\ref{eqn:a4})),
we obtain the following equation:
\begin{eqnarray}
\frac{dE}{dt} & = & m_3 \left\{
\frac{ R \dot{r} \cos(\phi - \psi) -
Rr\dot{\phi} \sin(\phi - \psi) - r \dot{r}}
{[R^2 + r^2 -2Rr\cos(\phi - \psi)]^{\frac{3}{2}}} \right.\nonumber \\
& & - \left.
\frac{ R \dot{r} \cos(\phi - \psi) - 
Rr\dot{\phi} \sin(\phi - \psi) + r \dot{r}}
{[R^2 + r^2 +2Rr\cos(\phi - \psi)]^{\frac{3}{2}}}\right\}~~ \label{eqn:a14}.
\end{eqnarray}

Substituting equations (\ref{eqn:a8}),(\ref{eqn:a9}) and
(\ref{eqn:a10}) into equation (\ref{eqn:a14}), 
and expanding with
respect to $a/R$, we obtain
\begin{eqnarray}
\frac{dE}{dt} & = & \frac{m_3}{R^3}
\left\{-2r\dot{r} + \frac{6r\cos(\phi - \psi)}{R}\left[
R\dot{r}\cos(\phi - \psi) - Rr\dot{\phi}\sin(\phi - \psi)\right]\right\}.
\label{eqn:diffe}
\end{eqnarray}

Assuming that the change in the position of particle 3 is negligible,
we can integrate equation (\ref{eqn:diffe}) along the unperturbed
orbit of the inner binary.
Integrating it over one period of the inner binary,
we obtain:
\begin{eqnarray}
\Delta E & = & 0. \label{eqn:e-varie}
\end{eqnarray}
This result is usually called as adiabatic invariance.

\subsection{Angular Momentum and Eccentricity}

The change in the specific angular momentum of the inner binary is
\begin{eqnarray}
A & = & 2~~\sum_{i=1,2}|{\bf r}_i \times {\bf v}_i|,\\
\delta A & = & 2~~\sum_{i=1,2}|\delta {\bf r}_i \times {\bf v}_i| +
|{\bf r}_i \times \delta {\bf v}_i|
\end{eqnarray}
Again we neglect the variation in $R$ as we did in the previous section
and replace $\delta {\bf v}$ with ${\bf a}\delta t$, to find
\begin{eqnarray}
\delta A & = & 2~~m_3\frac{\frac{1}{2}a(1-e^2)}
{(1+e\cos\phi)} 
\left\{ 
\frac{\left[(\cos \phi, \sin \phi)\times 
(R\cos\psi - r\cos\phi, R\sin\psi - r\sin\phi )\right]}
{[R^2+r^2-2Rr\cos(\phi-\psi)]^{\frac{3}{2}}} \right. \nonumber \\
 &  & ~~~ + ~~~ \left. \frac{\left[-(\cos \phi, \sin \phi)\times 
(R\cos\psi + r\cos\phi, R\sin\psi + r\sin\phi )\right]}
{[R^2+r^2+2Rr\cos(\phi-\psi)]^{\frac{3}{2}}} 
\right\}
\nonumber \\
& = & 2~~m_3\frac{\frac{1}{2}Ra(1-e^2)}
{(1+e\cos\phi)[R^2+r^2-2Rr\cos(\phi-\psi)]^{\frac{3}{2}}}
\sin(\psi - \phi).??
\end{eqnarray}
Expanding equation {\bf ??} with respect to $a/R$, and neglecting
higher terms, we obtain:
\begin{eqnarray}
\frac{dA}{dt} & = & 2~~m_3\frac{a(1-e^2)}{2R^2}
\left[\frac{\sin(\psi - \phi)}{1+e\cos \phi} \right] 
\left[\frac{6a(1-e^2)}{2R} \frac{\cos(\phi - \psi)}{1 + e \cos\phi}
\right].
\end{eqnarray}

Assuming that the position of particle 3 does not change during one
period of the inner binary (i.e. $R$ and $\psi $ are constant),
we integrate equation {\bf ??} over one period of the unperturbed motion of
the inner binary:
\begin{eqnarray}
\Delta A & = & \frac{3 m_3~~
a^{\frac{7}{2}}(1-e^2)^{\frac{7}{2}}}
{R^3} \sin \psi \cos \psi B(e) .\label{eqn:a-varie}
\end{eqnarray}

From the variation of energy and angular momentum, 
we obtain the variation of eccentricity for one period as follows:
\begin{eqnarray}
\Delta e & = & - \frac{(1-e^2)^{\frac{1}{2}}}
{ e a^{\frac{1}{2}}} \Delta A + \frac{a(1-e^2)}{e}\Delta E \nonumber \\
& = & - \frac{m_3 a^3}{R^3~e}
\left\{3(1-e^2)^4 B(e) \right\}
\sin \psi \cos \psi ,\label{eqn:ecc-varie}
\end{eqnarray}
where the function $B(e)$ is defined as
\begin{eqnarray}
B(e) \equiv \int_0^{2\pi} \frac{2\cos^2\phi - 1}{(1+e\cos \phi)^4} d\phi.
\end{eqnarray}
For our numerical experiments, the value of $B(e)$
is $B(0.9) = 4260$.

\subsection{Secular Variation}

Integrating equations (\ref{eqn:e-varie}), (\ref{eqn:a-varie}) and
(\ref{eqn:ecc-varie}), the variations of the specific angular momentum
and the eccentricity of the inner binary are 
%{\bf (we haven't used $E_0$ elsewhere---perhaps you should do so
%systematically throughout --- Steve)}
\begin{eqnarray}
%
%%% Energy
%
\frac{\Delta E(t)}{E} & = & 0.0,
\label{eqn:de}\\
%
%%% Angular Momentum
%
\Delta A(t) & = & \frac{3m_3}{8\pi}\left(\frac{P_{out}}{P_{in}}\right)
\frac{a^{\frac{7}{2}}(1-e^2)^{\frac{7}{2}}}{R^3}
B(e)\left[1-\cos (2\Omega_0 t)\right]  ,
\label{eqn:da}\\
%
%%% Eccentricity
%
\Delta e(t) & = & \frac{3~m_3}{8\pi}\left(\frac{P_{out}}{P_{int}}\right)
\frac{a^3(1-e^2)^4}{R^3 e} B(e)
\left[-1+\cos (2\Omega_0 t)\right].
\label{eqn:decc}
\end{eqnarray}
Here $P_{in}$ is the unperturbed period of the inner binary.

For the case of our numerical experiments 
(i.e. $e = 0.9$), equations (\ref{eqn:de})--(\ref{eqn:decc})
yield
\begin{eqnarray}
\frac{\Delta E(t)}{E} & \cong & 0.0,\\
\Delta A(t) & \cong & 0.46\cdot 10^{-3}(1-\cos 2\Omega_0 t) ,\\
\Delta e(t) & \cong & 0.23\cdot 10^{-3}(-1+\cos 2\Omega_0 t) .
\end{eqnarray}

\newpage

%
% References
%

\newpage

% 
% Figure captions
%
\begin{figure}

\figcaption[Funato.fig1.ps]{
Effect of stabilizing parameter $\alpha $ on time variation
of error in binding energy of a binary with eccentricity $e = 0.9$.
Figures (a) through (f) correspond to the cases of $\alpha = $ 0.1,
0.2, 0.3, 0.4, 0.5, and 0.6, respectively.
\label{fig0}
}

\figcaption{
Time evolution of errors for the case $e = 0.999999$.
Unit of time is the period of the binary.
2a). Relative error in the specific energy of the binary.
2b). Error in the angular momentum.
2c). Error in the eccentricity.
\label{fig1}
}

\figcaption[Funato.fig3.ps]{
Time evolution of relative error in binding error for the
simple binary case.  Unit of the horizontal axis is the period of the
binary in KS coordinates.  Solid, dashed and long-dashed lines
correspond to the cases of symmetrized, ``plain'' and stabilized case,
respectively. 
\label{fig2}
}

\figcaption[Funato.fig4.ps]{
Same as Figure 3 but for the angular momentum.  Top and bottom
are the same figures except for the scale of vertical axis.
\label{fig3}
}
\figcaption[Funato.fig5.ps]{
Same as Figure 3 but for the eccentricity.
\label{fig4}}

\figcaption[Funato.fig6.ps]{
Time evolution of the error of total energy for the
perturbed binary case.
The curves are same as those in Figure 3
The unit of horizontal axis are the period of the inner binary
in its KS coordinates.
\label{fig5}
}

\figcaption[Funato.fig7.ps]{
Same as Figure 6 but for the angular momentum.
\label{fig6}}

\figcaption[Funato.fig8.ps]{
Time evolution in
the specific energy of the inner binary for the perturbed binary
case. The curves and 
and the unit of horizontal axis
are same as those in Figure 6.
\label{fig7}
}

\figcaption[Funato.fig9.ps]{
Same as Figure 8 but for the specific angular momentum of
the inner binary.
\label{fig8}
}

\figcaption[Funato.fig10.ps]{
Same as Figure 8 but for the eccentricity of the inner
binary.
\label{fig9}
}

\figcaption{
The evolution of the inner binary for the symmetrized case.
Unit of time is the period of the inner binary.
Solid and dashed curves
correspond to the numerical experiment and theoretical,
respectively.
11a). Angular momentum.
11b). Eccentricity.
\label{fig10}
}

\figcaption[Funato.fig12.ps]{
Coordinates of the hierarchical triple.
\label{figa1}
}

\end{figure}

%
% Tables
%

\newpage

%document style
%\documentstyle[12pt,../style/aj_pt4]{article}

\pagestyle{empty}

\tablenum{1}

%\begin{document}

\begin{table}[h]

\caption{
Initial orbit parameters for the simple binary case.
\label{table:1}} 

\begin{center}
\begin{tabular}{crrrrrr}
\tableline
\tableline
& $m_1$ & $m_2$ & $a$ & $h$ & $e$ & period[2$\pi$] \\
\tableline
& 0.5 & 0.5 & 1.0 & -0.5 & 0.9e+0 & 1 \\
\tableline
\tableline
\end{tabular}

\end{center}
\end{table}

\tablenum{2}

\begin{table}[h]

\caption{
Initial orbit parameters for the hierarchical triplet case.
\label{table:2}}

\begin{center}
\begin{tabular}{crrrrrr}
\tableline
\tableline
& $m_1$ & $m_2$ & $a$ & $h$ & $e$ & period[$2\pi$] \\
\tableline
inner& 0.5 & 0.5 & 1.0 & -0.5 & 0.9e+0 & 1 \\
\tableline
outer& 1.0 & 0.01 & 10.1 & -0.05 & 0.0e+0 & 32 \\
\tableline
\tableline
\end{tabular}

\end{center}

\end{table}

\end{document}